\documentclass[twocolumn]{aastex701}
\usepackage{graphics,epsf}
\usepackage{amsmath}                
\usepackage{amsfonts}               
\usepackage{amssymb}                
\usepackage{epsfig}                 
\usepackage{appendix}
\usepackage{graphicx}
\usepackage{float}
\usepackage{color}
\usepackage{multirow}
\usepackage{colortbl}
\usepackage[para,online,flushleft]{threeparttable}

\hypersetup{citecolor=blue, 
            linkcolor=red, 
            menucolor=blue, 
            urlcolor=blue}  

\newcommand{\m}{{~\rm m}}

\newcommand{\s}{{~\rm s}}
\newcommand{\h}{{~\rm h}}

\newcommand{\g}{{~\rm g}}
\newcommand{\G}{{~\rm G}}
\newcommand{\K}{{~\rm K}}

\begin{document}

\title{Identifying a circum-jet southern ring counterpart to the northern jet of the Crab Nebula}


\author[0000-0003-0375-8987]{Noam Soker}
\affiliation{Department of Physics, Technion - Israel Institute of Technology, Haifa, 3200003, Israel;  soker@technion.ac.il}
\email{soker@technion.ac.il}

\begin{abstract}
I analyze images of the Crab Nebula core-collapse supernova (CCSN) remnant in light of recent three-dimensional hydrodynamical simulations of the jittering-jets explosion mechanism (JJEM) and identify a southern ring opposite to the northern jet, which I attribute to a counterjet. The Crab Nebula is known for its point-symmetric morphology of seven pairs of bays and a pair of two filaments, but no pairs of two jets or their direct outcomes, like ears and rings, have been identified. I identify a ring in visible and infrared images of the Crab Nebula opposite to the prominent northern jet. Recent hydrodynamical simulations of the JJEM show that jets that explode CCSNe can form such circum-jet rings. I, therefore, attribute the shaping of the southern ring to a southern jet, a counter jet to the northern jet, both of which participated in the explosion of the Crab Nebula in the framework of the JJEM.   
\end{abstract}
 

\section{Introduction} 
\label{sec:intro}

The morphologies of core-collapse supernova (CCSN) remnants (CCSNRs), particularly point-symmetric morphologies, are the only observations capable of robustly distinguishing between the two intensively studied alternative theoretical explosion mechanisms of CCSNe: the neutrino-driven mechanism (e.g., \citealt{Mezzacappa2026}) and the jittering-jets explosion mechanism (JJEM; e.g., \citealt{Soker2025Learning}). 

The historic and iconic Crab Nebula is one of about twenty point-symmetric CCSNRs. \cite{ShishkinSoker2025Crab} identified its point symmetric morphology in JWST IR observations: seven pairs of `bays', and one pair of filaments (\citealt{Blairetal2026} rediscovered this pair). \cite{Temimetal2024} identified most of these bays, but not the point-symmetric morphology. Although the point-symmetric morphology is predicted by the JJEM, the bays and the filaments are not necessarily along axes of jet pairs. 
\cite{GrichenerSoker2017} attributed the long axis of the Crab Nebula to a jet axis in the framework of the JJEM based on the southeast protrusion (an `ear'). But there is no ear on the other side. 

Recently, \cite{Dingetal2026} analyzed the northern jet of the Crab Nebula. They claim there is no counterjet and discuss seven alternative explanations of the jet, none of which include the JJEM. In this Research Note, I use recent three-dimensional (3D) hydrodynamical simulations of the JJEM and argue that there is a counterstructure, which I attribute to the northern jet's counterjet. 

\section{A southern ring} 
\label{sec:ARing}

\begin{figure*}
\begin{center}
\includegraphics[trim=0cm 15.0cm 0cm 0cm, clip, scale=0.85]{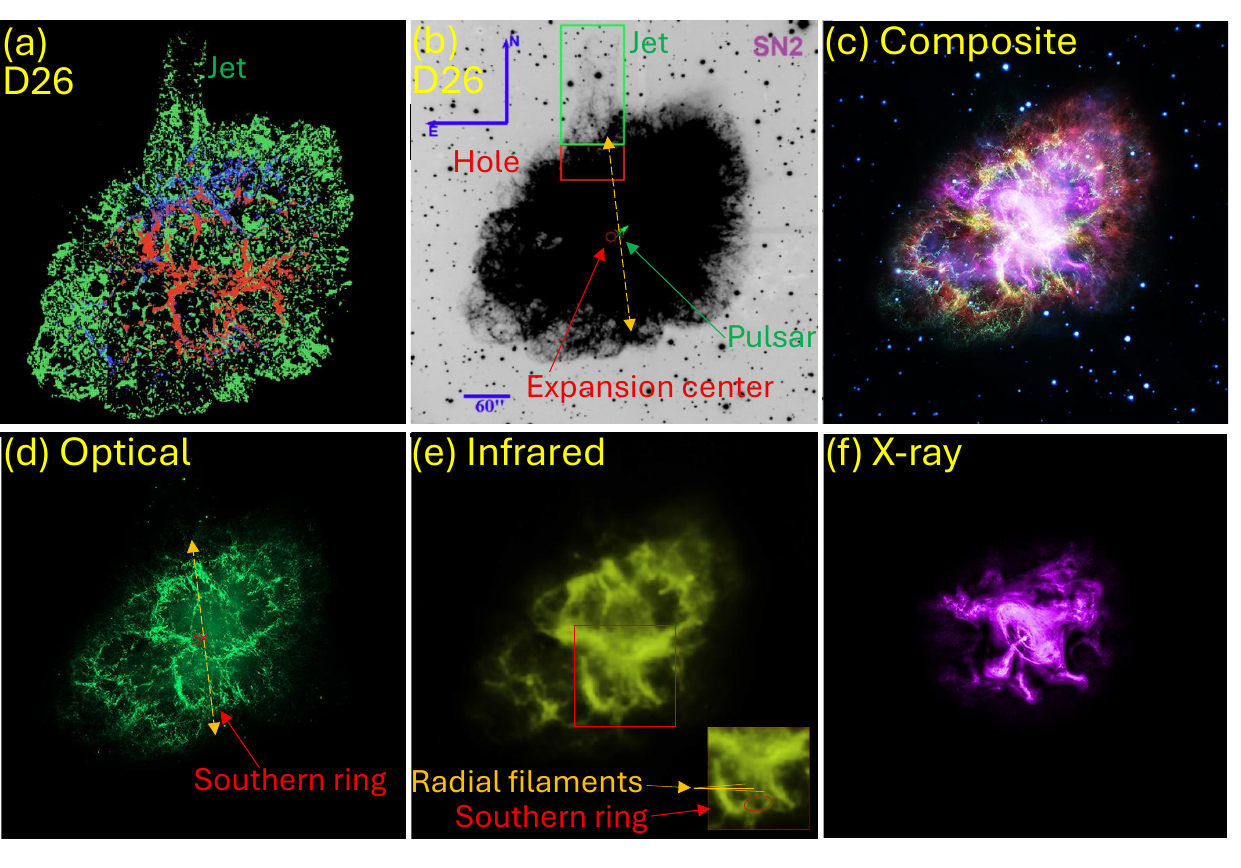} 
\caption{Images of the Crab Nebula with the new identification of the southern ring as a counterstructure to the northern jet.  
(a) An image adapted from \cite{Dingetal2026} presenting their 3-dimensional reconstruction of the Crab Nebula in [O \textsc{iii}] $\lambda \lambda$4959, 5007 (green), H$\beta$ (blue), and He I $\lambda$5876 (red). This image presents the northern jet structure as in \cite{Dingetal2026}. The jet was identified by \cite{vandenBergh1970}.  
(b) An optical figure adapted from \cite{Dingetal2026}. The green and red boxes indicating the northern jet region and the hole region, respectively, are from the original figure in \cite{Dingetal2026}. The circle at the center represents the expansion center determined by \cite{Nugent1998}, the ``X'' at the center is the location of the Crab pulsar (e.g., \citealt{Kaplanetal2008}), and the short arrow on the X indicates pulsar proper motion \citep{NgRomani2006}; all are marks from \cite{Dingetal2026}. I added only the double-sided orange arrow, copied from panel (d). 
Panels (c)-(f) are from the \href{https://chandra.harvard.edu/photo/2017/crab/}{Chandra Site} (\citealt{Dubneretal2017Crab}; Credits: X-ray: NASA/CXC/SAO; Optical: NASA/STScI; Infrared: NASA/JPL/Caltech; Radio: NSF/NRAO/VLA; Ultraviolet: ESA/XMM-Newton). They are all in the same scale as panel (b); panel (a) has a different scale. (c) A composite image: X-ray (Purple), Ultraviolet (Blue), Optical (Green), Infrared (Yellow-Green), and Radio (Red). (d) An optical image. I added the pulsar position and expansion center marks, identified the ring, and drew double-sided arrows to mark my proposed pair of jet axes (the arrow lengths have no meaning). (e) An infrared image where the inset shows the rings I identified. (f) An X-ray image to show that the pulsar-wind nebula is not correlated with the pair of jets.    
}
\label{fig:CrabNebula}
\end{center}
\end{figure*}

In a recent study, \cite{Dingetal2026} thoroughly analyzed the northern jet of the Crab Nebula; I present their image in panel (b) of Figure \ref{fig:CrabNebula}, and their 3D reconstruction in panel (a). The northern jet is not an active jet but rather a structure shaped by a jet, with bright, wavy radial filaments. Recent 3D hydrodynamical simulations show that jets can form such structures of a 'long ear' composed of radial filaments extending out from the main ejecta (e.g., \citealt{AkashiSoker2026a}); vortices and Kelvin-Helmholtz instability make the radial filaments wiggle.     

\cite{Dingetal2026} note that there is no counterjet in the south. Instead of a long ear with long radial filaments in the outer ejecta, as the northern jet is, I look for other structures that jets leave behind, like shorter radial filaments in the inner ejecta (e.g., \citealt{Braudoetal2026}) and circum-jet rings (e.g., \citealt{AkashiSoker2026a, AkashiSoker2026BG11}). I mark these features in panels (d) and (e) of Figure \ref{fig:CrabNebula}. I attribute the southern ring and the radial filaments to the northern jet's counterjet. Pairs with unequal opposite structural features are common in CCSNRs, mainly due to unequal pairs of jets in some pairs. 

The ratio of about 2 between the long and short axes of the ellipse of the southern ring suggests that the jet axis was at $\approx 30^\circ$ to the plane of the sky.

\section{Summary} 
\label{sec:Summary}

I identified a ring and radial filaments in an opposite direction to the northern jet of the Crab Nebula (Figure \ref{fig:CrabNebula}). In some cases, jets in simulations of the JJEM form a ring and filaments. Therefore, I attribute the ring and filaments in the Crab Nebula to a counterjet of the northern jet. I took the center of this pair of jets at the pulsar because I suggest that the neutron star launched this pair after it acquired its natal kick velocity. There are two reasons for this: (1) Late jets form more prominent rings \citep{AkashiSoker2026a}. (2) post-kick pairs of jets tend to have their axis at a large angle to the kick velocity (e.g., \citealt{BearSoker2023RNAAS}). The axis I argue for (the double-sided orange arrow in panel d) is at $54^\circ$ to the kick velocity, as marked. On the other hand, the long axis of the Crab Nebula and the kick velocity are almost aligned. The jets that form the elongated structure during the explosion could have imparted the kick to the pulsar (in what is termed the kick-BEAP mechanism; \citealt{Bearetal2025Puppis}). 

I consider the most likely explanation for the northern jet of the Crab Nebula to be that it is part of a pair of jets that participated in the explosion process within the framework of the JJEM.  

I encourage deep observations outside the main nebula, along the southern ring direction, to look for remnants of the southern jet, which I argue existed during the explosion. My findings add to the claim that the Crab Nebula is a point-symmetric CCSNR, as the JJEM predicts for many CCSNRs.  

\bibliography{BibReference}{}

@ARTICLE{BearSoker2023RNAAS,
       author = {{Bear}, Ealeal and {Soker}, Noam},
        title = "{The Jets and the Neutron Star Kick Velocity of the Supernova Remnant CTB 1}",
      journal = {Research Notes of the American Astronomical Society},
         year = 2023,
        month = dec,
       volume = {7},
       number = {12},
          eid = {266},
        pages = {266},
          doi = {10.3847/2515-5172/ad1392},
archivePrefix = {arXiv},
       eprint = {2312.02026},
 primaryClass = {astro-ph.HE},
       adsurl = {https://ui.adsabs.harvard.edu/abs/2023RNAAS...7..266B}
}

@ARTICLE{GrichenerSoker2017,
       author = {{Grichener}, Aldana and {Soker}, Noam},
        title = "{Core collapse supernova remnants with ears}",
      journal = {\mnras},
         year = 2017,
        month = jun,
       volume = {468},
       number = {1},
        pages = {1226-1235},
          doi = {10.1093/mnras/stx534},
archivePrefix = {arXiv},
       eprint = {1610.09647},
 primaryClass = {astro-ph.HE},
       adsurl = {https://ui.adsabs.harvard.edu/abs/2017MNRAS.468.1226G}
}

@ARTICLE{ShishkinSoker2025Crab,
       author = {{Shishkin}, Dmitry and {Soker}, Noam},
        title = "{Et tu, Brute?: The Crab Nebula also exploded by jittering jets}",
      journal = {arXiv e-prints},
         year = 2024,
        month = nov,
          eid = {arXiv:2411.07938},
        pages = {arXiv:2411.07938},
archivePrefix = {arXiv},
       eprint = {2411.07938},
 primaryClass = {astro-ph.HE},
       adsurl = {https://ui.adsabs.harvard.edu/abs/2024arXiv241107938S}
}

@ARTICLE{Soker2025Learning,
       author = {{Soker}, Noam},
        title = "{Learning from core-collapse supernova remnants on the explosion mechanism}",
      journal = {\na},
         year = 2025,
        month = dec,
       volume = {121},
          eid = {102453},
        pages = {102453},
          doi = {10.1016/j.newast.2025.102453},
archivePrefix = {arXiv},
       eprint = {2409.13657},
 primaryClass = {astro-ph.HE},
       adsurl = {https://ui.adsabs.harvard.edu/abs/2025NewA..12102453S}
}

@ARTICLE{Bearetal2025Puppis,
       author = {{Bear}, Ealeal and {Shishkin}, Dmitry and {Soker}, Noam},
        title = "{The Puppis A Supernova Remnant: An Early Jet-driven Neutron Star Kick followed by Jittering Jets}",
      journal = {Research in Astronomy and Astrophysics},
         year = 2025,
        month = apr,
       volume = {25},
       number = {4},
          eid = {045008},
        pages = {045008},
          doi = {10.1088/1674-4527/adc24e},
archivePrefix = {arXiv},
       eprint = {2409.11453},
 primaryClass = {astro-ph.HE},
       adsurl = {https://ui.adsabs.harvard.edu/abs/2025RAA....25d5008B}
}

@ARTICLE{AkashiSoker2026a,
       author = {{Akashi}, Muhammad and {Soker}, Noam},
        title = "{Simulating the jittering-jets explosion mechanism: circum-jet rings account for observed core-collapse supernova remnant morphologies}",
      journal = {arXiv e-prints},
         year = 2026,
        month = mar,
          eid = {arXiv:2603.29527},
        pages = {arXiv:2603.29527},
archivePrefix = {arXiv},
       eprint = {2603.29527},
 primaryClass = {astro-ph.HE},
       adsurl = {https://ui.adsabs.harvard.edu/abs/2026arXiv260329527A}
}

@ARTICLE{AkashiSoker2026BG11,
        author = {{Akashi}, Muhammad and {Soker}, Noam},
        title = "{Reproducing morphological features in the supernova remnant G11.2-0.3 by simulating jittering jets}",
      journal = {arXiv e-prints},
         year = 2026,
        month = may,
          eid = {arXiv:2605.12356},
        pages = {arXiv:2605.12356},
          doi = {10.48550/arXiv.2605.12356},
archivePrefix = {arXiv},
       eprint = {2605.12356},
 primaryClass = {astro-ph.HE},
       adsurl = {https://ui.adsabs.harvard.edu/abs/2026arXiv260512356A}
}

@ARTICLE{Braudoetal2026,
       author = {{Braudo}, Jessica and {Michaelis}, Amir and {Akashi}, Muhammad and {Soker}, Noam},
        title = "{Simulating observed point-symmetric core-collapse supernova morphologies with the jittering jets explosion mechanism}",
      journal = {arXiv e-prints},
         year = 2026,
        month = jun,
          eid = {arXiv:2606.14364},
        pages = {arXiv:2606.14364},
archivePrefix = {arXiv},
       eprint = {2606.14364},
 primaryClass = {astro-ph.HE},
       adsurl = {https://ui.adsabs.harvard.edu/abs/2026arXiv260614364B}
}

@ARTICLE{Mezzacappa2026,
       author = {{Mezzacappa}, Anthony},
        title = "{Core Collapse Supernova Modeling: The Next Ten Years}",
      journal = {arXiv e-prints},
         year = 2026,
        month = apr,
          eid = {arXiv:2604.24970},
        pages = {arXiv:2604.24970},
archivePrefix = {arXiv},
       eprint = {2604.24970},
 primaryClass = {astro-ph.HE},
       adsurl = {https://ui.adsabs.harvard.edu/abs/2026arXiv260424970M}
}

@ARTICLE{Dingetal2026,
       author = {{Ding}, Ziwei and {Milisavljevic}, Dan and {Martin}, Thomas and {Temim}, Tea and {Raymond}, John C. and {Mandal}, Soham and {Drissen}, Laurent},
        title = "{3D Kinematic Reconstruction of the Crab Nebula That Includes the Northern Ejecta `Jet'}",
      journal = {arXiv e-prints},
         year = 2026,
        month = jun,
          eid = {arXiv:2606.26231},
        pages = {arXiv:2606.26231},
          doi = {10.48550/arXiv.2606.26231},
archivePrefix = {arXiv},
       eprint = {2606.26231},
 primaryClass = {astro-ph.HE},
       adsurl = {https://ui.adsabs.harvard.edu/abs/2026arXiv260626231D}
}

@ARTICLE{Kaplanetal2008,
       author = {{Kaplan}, D.~L. and {Chatterjee}, S. and {Gaensler}, B.~M. and {Anderson}, J.},
        title = "{A Precise Proper Motion for the Crab Pulsar, and the Difficulty of Testing Spin-Kick Alignment for Young Neutron Stars}",
      journal = {\apj},
         year = 2008,
        month = apr,
       volume = {677},
       number = {2},
        pages = {1201-1215},
          doi = {10.1086/529026},
archivePrefix = {arXiv},
       eprint = {0801.1142},
 primaryClass = {astro-ph},
       adsurl = {https://ui.adsabs.harvard.edu/abs/2008ApJ...677.1201K}
}

@ARTICLE{NgRomani2006,
       author = {{Ng}, C.-Y. and {Romani}, Roger W.},
        title = "{Proper Motion of the Crab Pulsar Revisited}",
      journal = {\apj},
         year = 2006,
        month = jun,
       volume = {644},
       number = {1},
        pages = {445-450},
          doi = {10.1086/503315},
archivePrefix = {arXiv},
       eprint = {astro-ph/0602255},
 primaryClass = {astro-ph},
       adsurl = {https://ui.adsabs.harvard.edu/abs/2006ApJ...644..445N}
}

@ARTICLE{Nugent1998,
       author = {{Nugent}, Richard L.},
        title = "{New Measurements of the Expansion of the Crab Nebula}",
      journal = {\pasp},
         year = 1998,
        month = jul,
       volume = {110},
       number = {749},
        pages = {831-836},
          doi = {10.1086/316199},
       adsurl = {https://ui.adsabs.harvard.edu/abs/1998PASP..110..831N}
}

@ARTICLE{Temimetal2024,
       author = {{Temim}, Tea and {Laming}, J. Martin and {Kavanagh}, P.~J. and {Smith}, Nathan and {Slane}, Patrick and {Blair}, William P. and {De Looze}, Ilse and {Bucciantini}, Niccol{\`o} and {Jerkstrand}, Anders and {Gountanis}, Nicole Marcelina and {Sankrit}, Ravi and {Milisavljevic}, Dan and {Rest}, Armin and {Lyutikov}, Maxim and {DePasquale}, Joseph and {Martin}, Thomas and {Drissen}, Laurent and {Raymond}, John and {Fox}, Ori D. and {Modjaz}, Maryam and {Spitkovsky}, Anatoly and {Strolger}, Louis-Gregory},
        title = "{Dissecting the Crab Nebula with JWST: Pulsar Wind, Dusty Filaments, and Ni/Fe Abundance Constraints on the Explosion Mechanism}",
      journal = {\apjl},
         year = 2024,
        month = jun,
       volume = {968},
       number = {2},
          eid = {L18},
        pages = {L18},
          doi = {10.3847/2041-8213/ad50d1},
archivePrefix = {arXiv},
       eprint = {2406.00172},
 primaryClass = {astro-ph.HE},
       adsurl = {https://ui.adsabs.harvard.edu/abs/2024ApJ...968L..18T}
}

@ARTICLE{Blairetal2026,
       author = {{Blair}, William P. and {Sankrit}, Ravi and {Milisavljevic}, Dan and {Temim}, Tea and {Laming}, J. Martin and {Slane}, Patrick and {Ding}, Ziwei and {Martin}, Thomas},
        title = "{The Crab Nebula Revisited Using HST/WFC3}",
      journal = {\apj},
         year = 2026,
        month = jan,
       volume = {997},
       number = {1},
          eid = {81},
        pages = {81},
          doi = {10.3847/1538-4357/ae2adc},
archivePrefix = {arXiv},
       eprint = {2512.11103},
 primaryClass = {astro-ph.SR},
       adsurl = {https://ui.adsabs.harvard.edu/abs/2026ApJ...997...81B}
}
\bibliographystyle{aasjournal}

\end{document}